# Short Paper: Experimental Characterization of *In Vivo* Wireless Communication Channels

A. Fatih Demir[1], Qammer H. Abbasi[2], Z. Esat Ankarali[1],
Marwa Qaraqe[2], Erchin Serpedin[2], and Huseyin Arslan[1,3]

[1]Department of Electrical Engineering, University of South Florida, Tampa FL, USA
[2]Department of Electrical and Computer Engineering, Texas A&M University, USA/Qatar
[3] Department of Electrical and Electronics Engineering, Istanbul Medipol University, Istanbul, Turkey
Email: {afdemir, zekeriyya}@mail.usf.edu, {qammer.abbasi, marwa, serpedin}@tamu.edu, arslan@usf.edu

*Abstract*—*In vivo* wireless medical devices have a critical role in healthcare technologies due to their continuous health monitoring and noninvasive surgery capabilities. In order to fully exploit the potential of such devices, it is necessary to characterize the *in vivo* wireless communication channel which will help to build reliable and high-performance communication systems. This paper presents preliminary results of experimental characterization for this fascinating communications medium on a human cadaver and compares the results with numerical studies.

*Index Terms* — *In vivo* channel characterization, in/on-body communication, wireless body area networks (WBAN), wireless implantable medical devices.

## I. INTRODUCTION

*In vivo* wireless medical devices have the potential to play a vital role in future healthcare technologies. Such technologies include, but are not limited to pacemakers, implantable cardiac defibrillators (ICDs), internal drug delivery devices, neurostimulators, and wireless capsule endoscopes (WCEs). These devices provide continuous health monitoring and reduce the invasiveness of surgery with the integration of wearable devices. In order to fully exploit the capabilities of such devices, it is necessary to characterize and model the *in vivo* wireless communication channel (implant to implant and implant to an external device).

There exists a tremendous ongoing research on *in vivo* channel characterization in recent years. It is known that this medium's characteristics are subject-specific and strongly dependent on antenna location and frequency of operation. Although numerical characterization of *in vivo* wireless communication channel is being investigated intensely, there are relatively few studies for experimental characterization in the literature. This paper presents an experimental characterization of the *in vivo* wireless communication channel on human cadaver at 915 MHz. The preliminary results are presented and compared with our numerical study [1].

## II. EXPERIMENTAL SETUP

The ethical approval for this study is obtained from Istanbul Medipol University. In order to validate the simulation results [1], the experiment is conducted on a human cadaver with a similar setup. The human male torso area is investigated at 915 MHz by measuring channel response through a vector network analyzer (VNA) while using two antennas, one (*in vivo*) [2], and other a dipole antenna (*ex vivo*) as illustrated in Fig. 1. The *in vivo* antenna is placed at six locations inside the body around the heart, stomach, and intestine by a physician. The antennas are located in the same orientation, and all return loss values are less than -7dB.

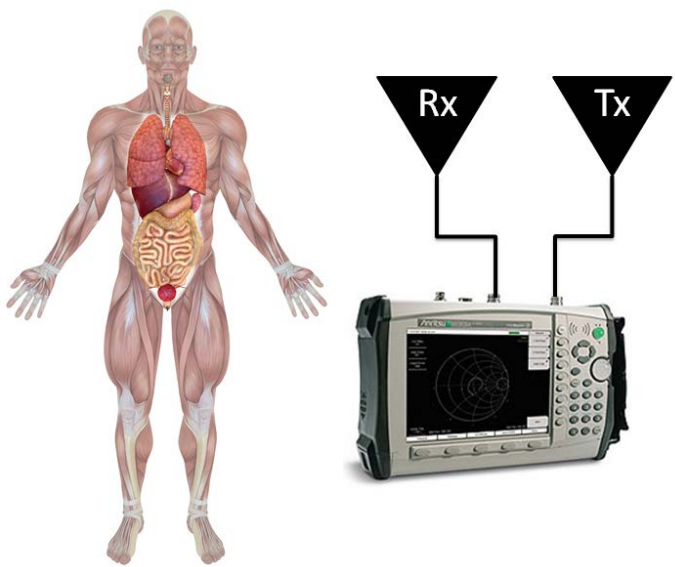

Fig. 1: Experimental setup for the *in vivo* channel.

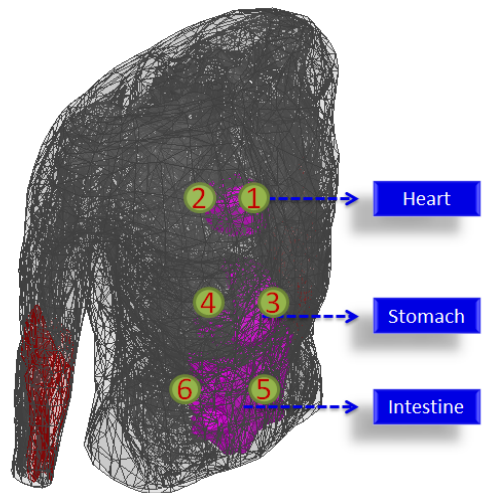

Fig. 2: Measurement locations on the human cadaver.

## III. RADIO CHANNEL CHARACTERIZATION

EM wave propagation inside the body is strongly related to the location of the antenna. Therefore, *in vivo* wireless channel characterization is mostly investigated for a specific part of the human body. Fig. 2 shows the antenna locations inside the body and Table I summarizes path loss values for these locations. The signal propagates through different organs and tissues that the path loss changes significantly for the locations at similar depth from the body surface.

Table I: Path loss values for selected *in vivo* locations.

| Location | *In Vivo* Depth | Path Loss |
|---|---|---|
| 01) Above Heart | 3 cm | 45.32 dB |
| 02) Below Heart | 8 cm | 55.61 dB |
| 03) Above Stomach | 5 cm | 48.19 dB |
| 04) Inside Stomach | 9 cm | 50.80 dB |
| 05) Above Intestine | 2 cm | 29.95 dB |
| 06) Below Intestine | 10 cm | 50.47 dB |

The channel modeling subgroup (Task Group 15.6), which worked on developing of the IEEE 802.15.6 standard, submitted their final report on body area network (BAN) channel models in November 2010 [3]. In this report, it is determined that Friis transmission equation can be used for in vivo scenarios by adding a random variation. In our previous study [1], the path loss is modeled as a function of depth by the following equation in dB:

$$PL(d) = PL_0 + m(d/d_0) + S \qquad (d \geq d_0)$$

where *d* stands for the depth from body surface in millimeters, $d_0$ denotes the reference depth (i.e. 10mm), $PL_0$ represents the intersection term in dB, *m* is the decay rate of received power, and *S* is the random shadowing parameter in dB, which presents a normal distribution for a fixed distance. The parameters for each anatomical body part are listed in the previous study. Table II compares the parameters which are obtained by numerical and experimental methods. The discrepancies should have occurred due to additional losses which were not considered in the simulation.

TABLE II: PARAMETERS FOR THE STATISTICAL PATH LOSS MODEL

| Environment | $PL_0$[dB] | m |
|---|---|---|
| Simulation | 23.04 | 2.28 |
| Experimental | 33.81 | 2.09 |

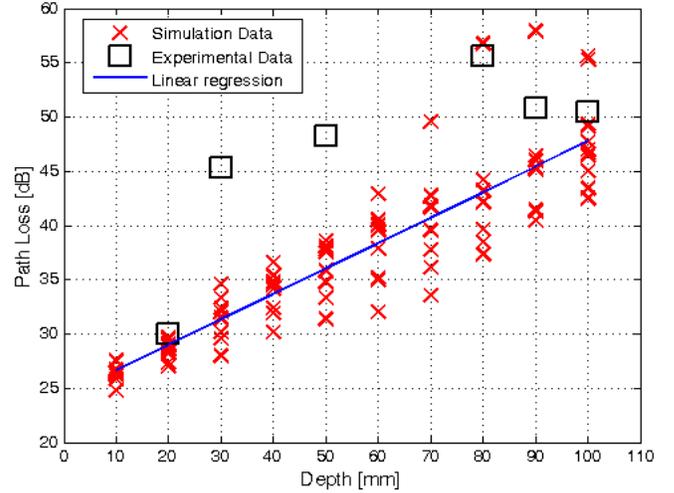

Fig. 4: Path loss versus depth from body surface.

## IV. CONCLUSION

To the best of our knowledge, this is the first study that experimentally investigates the *in vivo* wireless communication channel on a human cadaver. The numerical and experimental characterizations of the *in vivo* channel at 915 MHz are compared, and preliminary results are presented. The results emphasize the importance of the antenna location. However, further research needs to be performed for more *in vivo* locations, operational frequencies and well-designed *in vivo* antennas.

ACKNOWLEDGEMENT

This publication was made possible by NPRP grant # NPRP 6 - 415 - 3 - 111 from the Qatar National Research Fund (a member of Qatar Foundation). The authors are thankful to Istanbul Medipol University, School of Medicine for providing the human cadaver and their valuable medical assistance during the experiment.